\documentclass[prd,showpacs]{revtex4}
\begin{document}
\title{Finite temperature Casimir effect for massive scalars in a magnetic field}
\author{Andrea Erdas}
\email{aerdas@loyola.edu}
\author{Kevin P. Seltzer}
\affiliation{Department of Physics, Loyola University Maryland, 4501 North Charles Street,
Baltimore, Maryland 21210, USA}
%\date{August, 2013}
\begin {abstract} %CHANGE
The finite
temperature Casimir effect
for a charged, massive scalar field confined between very large, perfectly conducting parallel plates is studied
using the zeta function regularization technique. The scalar field satisfies Dirichlet boundary
conditions at the plates and a magnetic field perpendicular to the plates is present. Four equivalent expressions 
for the zeta function are obtained, which are exact to all orders in the magnetic 
field strength, temperature, scalar field mass, and plate distance. The zeta function is used to calculate 
the Helmholtz free energy of the scalar field and the Casimir pressure on the plates, in the case of high temperature, small plate distance,
strong magnetic field and large scalar mass. In all cases, simple analytic expressions of the zeta function,
free energy and pressure are obtained, which are very accurate and valid for practically all values of temperature, plate distance, magnetic field and mass.
\end {abstract}
\maketitle
%%%%%%%%%%%%%%%%%%%%%%%%%%%%%%%%%%%%%%%%%%%%%%%%%%%%%%%%%%%%%%%%%%%%%CHANGE
\section{Introduction}
\label{1}
In the Casimir effect, an attractive force is observed between perfectly
conducting and electrically neutral parallel plates in vacuum. Casimir's theoretical prediction of the effect was achieved
by calculating the attractive force between two neutral and parallel conducting plates caused by the electromagnetic field
quantum fluctuations in vacuum \cite{Casimir:1948dh}. 
A repulsive Casimir effect exists too, and was theoretically predicted by Boyer some time later when he showed that, in the 
case of a perfectly conducting sphere, the quantum fluctuations of the electromagnetic field produce a repulsive
force on the wall of the sphere \cite{Boyer:1968uf}. 
The first experimental proof of the Casimir force was obtained by Sparnaay \cite{Sparnaay:1958wg},
with many and more precise experimental observations reported since then. A comprehensive review 
of these experiments is presented in Refs. \cite{Bordag:2001qi,Bordag:2009zz}.

The interdisciplinary character of the Casimir effect is well known, 
since it is relevant not 
only in QED, but also in condensed matter physics, theories 
with compactiÞed extra dimensions, gravitation and cosmology,
mathematical physics, and nanotechnology and nanotubes. Therefore, a large effort has gone into studying this effect
and its generalization to quantum fields other than the electromagnetic field: fermions \cite{Johnson:1975,CougoPinto:2001ps} 
and especially scalar fields have been investigated extensively \cite{Bordag:2001qi}. 

Casimir forces are very sensitive to the quantum field boundary conditions on the plates and, in the case of scalar fields, the most frequently used
boundary conditions are Dirichlet and Neumann, while for fermion fields \cite{Chodos:1974je} or vector fields \cite{Ambjorn:1981xw} 
bag boundary conditions are used. Here we use the simplest boundary conditions, Dirichlet, to constrain a scalar field 
between two perfectly conducting parallel plates. 

Scalar fields, either massive or massless, appear everywhere in physics.
The Higgs field, responsible for spontaneous symmetry breaking in the Standard Model, 
is a massless scalar before the $SU(2)$ gauge symmetry is broken and a massive scalar after 
the symmetry is broken. Scalar fields are found within superstring theories as dilaton
fields, breaking the conformal symmetry of the string \cite{Brans:2005ra}. Scalar fields
are used to cause inflation, helping to solve the horizon problem and giving a
reason for the non-vanishing cosmological constant. Massless fields are used in this context
as inflatons, and massive ones (e.g. Higgs-like fields) are also used \cite{CervantesCota:1994zf}.
Scalar fields are 
used to explain Landau diamagnetism \cite{Biswas,Brito}, and in other areas of condensed matter physics. It has been shown that
the electromagnetic Casimir force between parallel plates is
obtained by simply doubling the Casimir force on the plates due to a massless scalar field,
where the factor of two accounts for the two polarization states of the photon. Therefore 
the Casimir force between parallel plates caused by a charged scalar field will be the 
same, apart from a multiplicative factor, as the force due to a charged vector field
such as the $W$-field or the gluon field.

The Casimir effect due to a charged scalar
field in the presence of a magnetic field has been studied in vacuum \cite{CougoPinto:1998td} and at finite temperature \cite{CougoPinto:1998jg}.
These authors use the Schwinger proper time method to
calculate the effective action, but are only able to obtain the free energy as an infinite sum of 
modified Bessel functions. In a recent paper \cite{Erdas:2013jga} we used a different 
method, the zeta function technique, to study the finite temperature Casimir effect for a massless and charged scalar
 %%%%%%%%%%%%%%%----PAGE 2----%%%%%%%%%%%%%%%%%%%%%%%%%%%%%%%%%%%%%
field in the presence of a magnetic field. We obtained 
simple analytic forms for the free energy and Casimir pressure, valid for practically all values of the three parameters involved.
In this paper we conduct a similar investigation of the Casimir effect at finite temperature, but we focus on a
massive and charged scalar field in the presence of a magnetic field. We obtain the corrections to the results of our recent paper
\cite{Erdas:2013jga} due to a ``light'' scalar mass and obtain new results for the cases of ``intermediate'' mass, where the scalar mass is 
larger than only some of the parameters, and ``large'' mass, where the scalar mass is the largest parameter. 

Casimir effect calculations generally follow 
Casimir's definition of the vacuum energy, which requires a regularization recipe for its implementation.
Many regularization  
techniques have been used for these calculations, such as the cutoff method 
in various piston configurations \cite{Cavalcanti:2003tw,Oikonomou:2009zr},
the world-line technique
\cite{Gies:2003cv}, the multiple-scattering method
\cite{Milton:2007wz,Milton:2011zv}, the zeta function technique 
\cite{Elizalde:2007du,Elizalde:2006iu,Elizalde:1988rh}, and others. As we stated above, in this paper we use 
the zeta function technique, a
regularization technique used also in the computation of effective actions \cite{dittrich,Erdas:1990gy}. We calculate
the free energy and Casimir pressure due to a massive scalar field, of mass $M$,
confined between two very large, perfectly conducting parallel plates, at a distance $a$ from each other.
The scalar field satisfies Dirichlet boundary conditions on the plates and our system is in thermal equilibrium 
with a heat reservoir at finite temperature $T$. We use the imaginary time 
formalism of finite temperature field theory, which is suitable for a system in thermal equilibrium. 
A uniform magnetic field $\vec B$ is present in the region between the plates and is perpendicular
to the plates.

In Sec. \ref{2}, we present four equivalent expressions of the zeta function for this system, exact to all orders in $B$, $T$, $M$, and $a$, and
obtain simple analytic expressions for the zeta function in the case of 
high temperature, small plate distance, strong magnetic field and large scalar mass.
In Sec. \ref{3} we use this zeta function to calculate the Helmholtz free energy 
of the scalar field and the pressure on the plates, and we obtain simple analytic expressions for these quantities
in the case of high temperature, small plate distance, strong magnetic field and large mass.
We discuss our results in Sec. \ref{4}. 
%%%%%%%%%%%%%%%%%%%%%%%%%%%%%%%%%%%%%%%%%%%%%%%%%%%%%%%%%%%%%%%%%%%%%
\section{ Zeta function evaluation}
\label{2}

We investigate a scalar field $\phi(x,\tau)$ of mass $M$ and charge $e$ in three-dimensional space and Euclidean time $\tau$,
confined by two large, square, perfectly conducting parallel plates  perpendicular to the $z$ axis and located at $z=0$ and $z=a$. We impose 
Dirichlet boundary conditions that constrain the scalar field to vanish at the plates
\begin{equation}
\phi(x,y,0,\tau)=\phi(x,y,a,\tau)=0,
\label{bag}
\end{equation}
and use finite temperature field theory to take into account temperature effects on our system, 
which we assume to be in thermal equilibrium at temperature $T$.
The imaginary time formalism of finite temperature field theory allows only field configurations satisfying the following
boundary conditions
\begin{equation}
\phi(x,y,z,\tau)=\phi(x,y,z,\tau+\beta),
\label{periodic}
\end{equation}
for any $\tau$, where $\beta=1/T$ is the periodic length in the Euclidean time axis.
In the slab region there is also a uniform magnetic field pointing in the $z$ direction, ${\vec B}=(0,0,B)$, and 
therefore the charged scalar field interacts with ${\vec B}$.

The Helmholtz free energy $F$ for the scalar field is
$$F=\beta^{-1}\log \,\det \left(D_{\rm E}|{\cal F}_a\right),$$
where the symbol ${\cal F}_a$ indicates the set of functions satisfying boundary conditions 
(\ref{bag}) and (\ref{periodic}), and the operator $D_{\rm E}$ is defined as:
$$D_{\rm E} = -\partial^2_\tau+p^2_z+({\vec p} -e{\vec A})^2_\perp+M^2,$$
where the subscript E indicates Euclidean time, $\vec A$ is the electromagnetic vector potential, and we use 
the notation ${\vec p}_\perp=(p_x,p_y,0)$.

The zeta function technique allows us to evaluate $F$ using the eigenvalues of $D_{\rm E}$ .
The Dirichlet boundary conditions (\ref{bag}) are satisfied 
if the $z$ component of the momentum is only allowed to take the values 
$$p_z={\pi\over a}n,$$
where $n = 0, 1, 2, 3,...$. The eigenvalues of the operator 
$({\vec p} -e{\vec A})^2_\perp$ are the Landau levels
$$2eB\left(l+{1\over 2}\right),$$
%%%%%%%%%%%%%%%----PAGE 3----%%%%%%%%%%%%%%%%%%%%%%%%%%%%%%%%%%%%%
with $l = 0, 1, 2, 3,...$, and therefore  the eigenvalues of $D_{\rm E}$ whose eigenfunctions satisfy (\ref{bag}) and (\ref{periodic}) are:
\begin{equation}
{\pi^2\over a^2}n^2+{4\pi^2\over\beta^2}m^2+eB\left(2l+1\right)+M^2,
\label{eigenvalues1}
\end{equation}
where $n,l = 0, 1, 2, 3,...$ and $m = 0, \pm 1, \pm 2, \pm 3,...$.
We use this set of eigenvalues to construct
the zeta function  of the operator $D_{\rm E}$
\begin{equation}
\zeta(s)=L^2  \sum_{n=0}^\infty \sum_{m=-\infty}^\infty\left(
{eB \over 2\pi}\right)\mu^{2s} \sum_{l=0}^\infty\left[
{\pi^2\over a^2}n^2+{4\pi^2\over\beta^2}m^2+eB\left(2l+1\right)+M^2
\right]^{-s},
\label{zetapm}
\end{equation}
where $L^2$ is the area of the plates, the factor $eB/ 2\pi$ takes into account the degeneracy per unit 
area of the Landau levels and the arbitrary parameter $\mu$ with the dimension of a mass has been introduced 
to keep $\zeta(s)$ dimensionless for all values of $s$.
Once we obtain $\zeta(s)$, we use the zeta function technique
and easily find the free energy by taking a simple derivative
\begin{equation}
F=-\beta^{-1}\zeta'(0).
\label{Fandzeta}
\end{equation}

The following identities
$$
x^{-s}={1\over \Gamma(s)}\int_0^\infty dt\, t^{s-1}e^{-xt},
$$
\begin{equation}
\sum_{l=0}^\infty e^{-(2l+1)x}={1\over 2 \sinh x},
\label{sinh}
\end{equation}
where $\Gamma (s)$ is the Euler gamma function, allow us to rewrite $\zeta(s)$ as
\begin{equation}
\zeta(s)= {L^2 \mu^{2s}\over 4\pi\Gamma(s)}
\int_0^\infty dt \, t^{s-2} e^{-M^2t} {eBt \over \sinh eBt}\left(\sum_{n=0}^\infty e^{-{\pi^2\over a^2}n^2 t}\right)
\left(\sum_{m=-\infty}^\infty e^{-{4\pi^2\over \beta^2}m^2 t}\right).
\label{zeta2}
\end{equation}
It is not possible to evaluate (\ref{zeta2}) in closed form for any value of the four quantities $B$, $M$, $a$ and $T$, but it 
is possible to obtain simple expressions of $\zeta(s)$  when one or some of these quantities are small or large. We will use these 
expressions of the zeta function in closed form to easily obtain the free energy using (\ref{Fandzeta}).

First we evaluate the zeta function in the high temperature limit, when $T\gg a^{-1}, M, \sqrt{eB}$, 
and apply Poisson resummation formula  \cite{Dittrich:1979ux} to 
the $n$ sum in (\ref{zeta2}), to obtain
\begin{equation}
\zeta(s)=a\left[\zeta_{M,B}(s)+\zeta_{M,B,a}(s)+\tilde{\zeta}_{M,B,T}(s) +
\zeta_{M,B,a,T}(s)\right],
\label{z}
\end{equation}
where
\begin{equation}
{\zeta}_{M,B}(s)=
{L^2\mu^{2s} \over 8\pi\Gamma(s)}
\int_0^\infty dt \, t^{s-2}\left({1\over{\sqrt {\pi t}}}+a^{-1}\right) e^{-M^2t}{eBt \over \sinh eBt},
\label{zB}
\end{equation}
\begin{equation}
\tilde{\zeta}_{M,B,T}(s)=
{L^2 \mu^{2s}\over 4\pi\Gamma(s)}\sum_{m=1}^\infty 
\int_0^\infty dt \, t^{s-2}\left({1\over{\sqrt {\pi t}}}+a^{-1}\right) e^{-M^2t}{eBt \over \sinh eBt} e^{-{4\pi^2m^2t}/\beta^2},
\label{ztBT}
\end{equation}
\begin{equation}
\zeta_{M,B,a}(s)=
{L^2 \mu^{2s}\over 4\pi^{3/2}\Gamma(s)}\sum_{n=1}^\infty 
\int_0^\infty dt \, t^{s-5/2} e^{-M^2t}{eBt \over \sinh eBt} e^{-{n^2a^2/t}},
\label{zBa}
\end{equation}
\begin{equation}
\zeta_{M,B,a,T}(s)=
{L^2\mu^{2s} \over 2\pi^{3/2}\Gamma(s)}\sum_{n=1}^\infty \sum_{m=1}^\infty 
\int_0^\infty dt \, t^{s-5/2} e^{-M^2t}{eBt \over \sinh eBt} e^{-({n^2a^2/t} + {4\pi^2m^2t}/\beta^2)}.
\label{zBaT}
\end{equation}

%%%%%%%%%%%%%%%----PAGE 4----%%%%%%%%%%%%%%%%%%%%%%%%%%%%%%%%%%%%%
We use (\ref{sinh}) to rewrite (\ref{zB}) as
\begin{equation}
{\zeta}_{M,B}(s)=
{L^2 (eB)^{{3\over 2}} \over 4\pi^{{3\over 2}}\Gamma(s)}\left({\mu^2\over eB}\right)^s
\int_0^\infty dt \, t^{s-{3\over 2}}e^{-zt}
\sum_{l=0}^\infty e^{-(2l+1)t},
\label{zB2}
\end{equation}
where $z=M^2/eB$ and we dropped the term proportional to $a^{-1}$, since it only contributes a constant independent of 
the plate distance to the free energy. After changing the integration variable from $t$ to $t\over (2l+1)+z$,
we find
$$
{\zeta}_{M,B}(s)=
{L^2 (eB)^{3\over 2} \over 4\pi^{{3\over 2}}}\left({\mu^2\over eB}\right)^s{\Gamma(s-{\scriptstyle{1\over 2}}) \over \Gamma(s)}
\left[\zeta_H(s-{\scriptstyle{1\over 2}},z)
-2^{{1\over 2}-s}\zeta_H(s-{\scriptstyle{1\over 2}},{\scriptstyle {z\over 2}})
\right],
$$
which is exact for all values of $s$, $B$ and $M$ and where 
$$
\zeta_H(s,z)=\sum_{l=0}^\infty (l+z)^{-s}
$$
is the Hurwitz zeta function. To
calculate the free energy, we only need to know $\zeta(s)$ for $s\rightarrow 0$. For small $s$ we find
\begin{equation}
x^{s}\zeta_H(s-{\scriptstyle{1\over 2}},z){\Gamma(s-{1\over 2})\over \Gamma(s)}=-2{\sqrt{\pi}}\zeta_H(-{\scriptstyle{1\over 2}},z)s +{\cal O}(s^2),
\label{lim1}
\end{equation}
and therefore
\begin{equation}
{\zeta}_{M,B}(s)=
{L^2 (eB)^{3\over 2} \over 2\pi}
\left[\sqrt{2}\zeta_H(-{\scriptstyle{1\over 2}},{\scriptstyle {z\over 2}})-
\zeta_H(-{\scriptstyle{1\over 2}},z)
\right]s,
\label{zB3}
\end{equation}
when $s$ is small.
In addition to Eq. (\ref{zB3}),  we have obtained two simpler expressions of ${\zeta}_{M,B}(s)$, one valid in the small mass
limit, $M^2\ll eB$, and the other in the large mass limit, $M^2\gg eB$. For 
$M^2\ll eB$ we take $e^{-zt}\approx 1-zt+{\cal O}(z^2)$ in (\ref{zB2}), integrate and find
\begin{equation}
{\zeta}_{M,B}(s)=
{L^2 (eB)^{3\over 2} \over 4\pi}\left({\mu^2\over eB}\right)^s
\left[
{\Gamma(s-{1\over 2})\over{\sqrt {\pi}\Gamma(s)}}\left(1-2^{{1\over 2}-s}\right)\zeta_R(s-{\scriptstyle{1\over 2}})
-z{\Gamma(s+{1\over 2})\over{\sqrt {\pi}\Gamma(s)}}\left(1-2^{-{1\over 2}-s}\right)\zeta_R(s+{\scriptstyle{1\over 2}})
+{\cal O}(z^2)\right],
\label{zB4}
\end{equation}
where $\zeta_R(s)$ is the Riemann zeta function of number theory. We use 
$$
x^{s}\zeta_R(s+{\scriptstyle{1\over 2}}){\Gamma(s+{1\over 2})\over \Gamma(s)}={\sqrt{\pi}}\zeta_R({\scriptstyle{1\over 2}})s +{\cal O}(s^2),
$$
and (\ref{lim1}), to obtain the small mass and small $s$ limit of ${\zeta}_{M,B}(s)$ 
\begin{equation}
{\zeta}_{M,B}(s)=
{L^2 (eB)^{3\over 2} \over 2\pi}(\sqrt 2 -1)
\left[
\zeta_R(-{\scriptstyle{1\over 2}})
-{M^2\over 2^{3\over 2}eB}\zeta_R({\scriptstyle{1\over 2}})
\right]s,
\label{zB51}
\end{equation}
where $\zeta_R (-{\scriptstyle{1\over 2}})=-0.2079$ and $\zeta_R ({\scriptstyle{1\over 2}})=-1.4603$.
When $M^2\gg eB$, we take
\begin{equation}
{eBt \over\sinh eBt}\approx 1-{1\over 6} (eBt)^2+{\cal O}(e^4B^4)
\label{smallB}
\end{equation}
inside Eq. (\ref{zB}), neglect again the term proportional to $a^{-1}$, integrate and find
$$
{\zeta}_{M,B}(s)=
{L^2 M^3 \over 8\pi^{3\over 2}\Gamma(s)}\left({\mu\over M}\right)^{2s}\left[{\Gamma(s-{\scriptstyle{3\over2}})}
-{e^2B^2\Gamma(s-{1\over2})\over 6 M^4}+{\cal O}\left({e^{4}B^4\over M^8}\right)\right],
$$
which, for $s\rightarrow 0$, becomes
\begin{equation}
{\zeta}_{M,B}(s)={L^2 M^3 \over 6\pi}\left(1
-{e^2B^2\over8M^4}
\right)s.
\label{zB6}
\end{equation}

%%%%%%%%%%%%%%%----PAGE 5----%%%%%%%%%%%%%%%%%%%%%%%%%%%%%%%%%%%%%
Next we evaluate $\tilde{\zeta}_{M,B,T}(s)$ for $eB\ll 4\pi^2T^2+M^2$. We substitute (\ref{smallB}) into (\ref{ztBT}), neglect once more the 
term proportional to $a^{-1}$, integrate, and find
\begin{equation}
\tilde{\zeta}_{M,B,T}(s)={L^2 \mu^{2s} \over 4\pi^{3\over 2} \Gamma(s)} 
\left[
{\Gamma(s-{\scriptstyle{3\over2}})} E_1^{M^2}(s-{\scriptstyle{3\over2}};4\pi^2T^2)
-{e^2B^2\over 6}\Gamma(s+{\scriptstyle{1\over2}})E_1^{M^2}(s+{\scriptstyle{1\over2}};4\pi^2T^2)\right],
\label{ztBT2}
\end{equation}
where we use Epstein functions 
\cite{Santos:1999yj,Elizalde:1988rh,Kirsten:1994yp}
which, for any positive integer $N$, are defined as
$$
E_N^{M^2}\left( s; \,a_1, a_2,..., a_N\right)=\sum_{n_1=1}^\infty\sum_{n_2=1}^\infty....\sum_{n_N=1}^\infty
{1\over (a_1n_1^2 + a_2 n_2^2 +....+ a_Nn_N^2+M^2)^s}.
$$
Eq. (\ref{ztBT2}) is valid for any value of $T$ and $M$, as long as $T\gg \sqrt{eB}$.
Since $T\gg M$, a simple analytic expression can be obtained for $\tilde{\zeta}_{M,B,T}(s)$, not involving Epstein functions
$$\tilde{\zeta}_{M,B,T}(s)={2\pi^{3\over 2}L^2T^3  \over  \Gamma(s)} \left({\mu\over 2\pi T}\right)^{2s}
\left[{\Gamma(s-{\scriptstyle{3\over2}})\zeta_R(2s-3)}
-{M^2\over 4\pi^{2}T^2}\Gamma(s-{\scriptstyle{1\over2}})\zeta_R(2s-1)
-{e^2B^2\over 96\pi^{4}T^4}\Gamma(s+{\scriptstyle{1\over2}})\zeta_R(2s+1)\right]
$$
which, for $s\rightarrow 0$, becomes
\begin{equation}
\tilde{\zeta}_{M,B,T}(s)={L^2T^3} 
\left[{\pi^2\over 45}-{M^2\over 12 T^2}
-{e^2B^2\over 48\pi^2T^4}\left(\ln {\mu\over 4\pi T}+\gamma_E+{1\over 2s}\right)
\right]s,
\label{ztBT3}
\end{equation}
where $\gamma_E = 0.5772$ is the Euler Mascheroni constant.

In the high temperature limit, it could happen that $M\gg \sqrt{eB}, a^{-1}$, or $\sqrt{eB}\gg M, a^{-1}$, or
 $a^{-1}\gg M, \sqrt{eB}$, and therefore 
we need to evaluate ${\zeta}_{M,B,a}(s)$ for each of these three different possibilities. When $M\gg \sqrt{eB}, a^{-1}$ we change
the integration variable from $t$ to $t n a\over M$ in Eq. (\ref{zBa}) and find
$$
{\zeta}_{M,B,a}(s) =
{L^2\mu^{2s} \over 4\pi^{3/2}\Gamma(s)}\sum_{n=1}^\infty  \left({na\over M}\right)^{s-1/2}
\!\!\!\int_0^\infty dt \, t^{s-5/2} {eBt \over \sinh ({eBtna\over M})} e^{-naM(t+t^{-1})}.
$$
Since $aM \gg1$, only the term with $n=1$ contributes significantly to the sum and, 
using the saddle point method, we evaluate the integral and find
$$
{\zeta}_{M,B,a}(s) =
{L^2 eB \over 4\pi a\Gamma(s)} \left({a\mu^2\over M}\right)^{s}
{e^{-2aM}\over\sinh ({eBa\over M})},
$$
which, for small $s$, becomes
 \begin{equation}
{\zeta}_{M,B,a}(s) =
{L^2 eB \over 4\pi a} {e^{-2aM}\over\sinh ({eBa\over M})}s.
\label{zBa2}
\end{equation}
When $\sqrt{eB}\gg M, a^{-1}$ we use (\ref{sinh}) into (\ref{zBa}) and change
the integration variable from $t$ to $t n a\over \sqrt{M^2+(2l+1)eB}$, to find
$$
{\zeta}_{M,B,a}(s) =
{L^2eB\mu^{2s} \over 2\pi^{3/2}\Gamma(s)}\sum_{n=1}^\infty\sum_{l=1}^\infty  \left({na\over \sqrt{M^2+(2l+1)eB}}\right)^{s-1/2}
\!\!\!\int_0^\infty dt \, t^{s-3/2} e^{-na\sqrt{M^2+(2l+1)eB}(t+t^{-1})}.
$$
Since $a\sqrt{eB}\gg 1$, only the term with $n=1$ and $l=0$ contributes significantly to the double sum and, 
using the saddle point method to evaluate the integral, we find
$$
{\zeta}_{M,B,a}(s) =
{L^2 eB \over 2\pi a\Gamma(s)} \left({a\mu^2\over \sqrt{M^2+eB}}\right)^{s}
e^{-2a\sqrt{M^2+eB}},
$$
and therefore, for small $s$
\begin{equation}
{\zeta}_{M,B,a}(s) =
{L^2 eB \over 2\pi a} 
e^{-2a\sqrt{M^2+eB}}s.
\label{zBa3}
\end{equation}
%%%%%%%%%%%%%%%----PAGE 6----%%%%%%%%%%%%%%%%%%%%%%%%%%%%%%%%%%%%%
When $a^{-1}\gg M, \sqrt{eB}$, we use $e^{-M^2t}\simeq 1- M^2t$ and (\ref{smallB}), integrate, and find
 $${\zeta}_{M,B,a}(s) =
{L^2 (\mu a)^{2s}\over 4\pi^{3\over 2}a^3\Gamma(s)}
\left[\Gamma({\scriptstyle{3\over 2}}-s)\zeta_R(3-2s)-M^2a^2\Gamma({\scriptstyle{1\over 2}}-s)\zeta_R(1-2s)-
{e^2B^2 a^4\over 6}\Gamma(-{\scriptstyle{1\over 2}}-s)\zeta_R(-1-2s)\right],
$$
which, in the small $s$ limit, becomes
\begin{equation}
{\zeta}_{M,B,a}(s) =
{L^2  \over 8\pi a^3} 
\left[\zeta_R(3)+M^2a^2\left(2\ln 2\mu a+{1\over s}\right)-{e^2B^2a^4\over18}\right]s,
\label{zBa4}
\end{equation}
where $\zeta_R(3)=1.2021$.

Last we evaluate $\zeta_{M,B,a,T}(s)$. After changing 
the integration variable from $t$ to $t n a\over\sqrt{ 4\pi^2 m^2T^2+M^2}$ in (\ref{zBaT}), we obtain
$$
\zeta_{M,B,a,T}(s)=
{L^2\mu^{2s} \over 2\pi^{3/2}\Gamma(s)}\sum_{n=1}^\infty \sum_{m=1}^\infty \left({na\over \sqrt{ 4\pi^2 m^2T^2+M^2}}\right)^{s-1/2}
\!\!\!\int_0^\infty dt \, t^{s-{5\over 2}} {eBt \over \sinh ({eBtna\over \sqrt{ 4\pi^2 m^2T^2+M^2}})} e^{-na\sqrt{ 4\pi^2 m^2T^2+M^2}(t+t^{-1})}.
$$
Since $aT \gg1$, only the term with $n=m=1$ contributes significantly to the double sum and, 
using the saddle point method, we evaluate the integral for $eB\ll 4\pi^2T^2+M^2$ to obtain
$$
\zeta_{M,B,a,T}(s)=
{L^2 eB \over 2\pi a\Gamma(s)} \left({a\mu^2\over \sqrt{ 4\pi^2 T^2+M^2}}\right)^{s}
{e^{-2a\sqrt{ 4\pi^2 T^2+M^2}}\over\sinh ({eBa\over \sqrt{ 4\pi^2 T^2+M^2}})},
$$
which, for small $s$, is
\begin{equation}
\zeta_{M,B,a,T}(s)=
{L^2 eB \over 2\pi a }
{e^{-2a\sqrt{ 4\pi^2 T^2+M^2}}\over\sinh ({eBa\over \sqrt{ 4\pi^2 T^2+M^2}})}s.
\label{zBaT2}
\end{equation}

We add (\ref{zB6}), (\ref{ztBT3}), (\ref{zBa2}) and  (\ref{zBaT2}), and find the high temperature and small $s$ limit of $\zeta(s)$, when
$T\gg M\gg \sqrt{eB}, a^{-1}$
\begin{equation}
\zeta(s)={L^2a} \left[
{\pi^2T^3\over 45}-{M^2T\over 12}+{M^3 \over 6\pi}
-{e^2B^2\over 48\pi^2T}\left({\pi T\over M}+\ln {\mu\over 4\pi T}+\gamma_E\right)
\right]s
+{ L^2 eB \over 4\pi }\left[ 
{e^{-2aM}\over\sinh ({eBa\over M})}+
{2e^{-2a\sqrt{ 4\pi^2 T^2+M^2}}\over\sinh ({eBa\over \sqrt{ 4\pi^2 T^2+M^2}})}
\right]s,
\label{z3}
\end{equation}
where we dropped a term independent of $s$ that does not contribute to the free energy.
We add (\ref{zB51}), (\ref{ztBT3}), (\ref{zBa3}) and  (\ref{zBaT2}), and find $\zeta(s)$ when
$T\gg \sqrt{eB}\gg M, a^{-1}$
\begin{eqnarray}
\zeta(s)&=&{L^2a} \left[
{\pi^2T^3\over 45}+{ (eB)^{3\over 2} \over 2\pi}(\sqrt 2 -1)
\zeta_R(-{\scriptstyle{1\over 2}})-{M^2T\over 12}-{\sqrt{eB}M^2\over 4\pi}\left(1-{1\over\sqrt 2}\right)\zeta_R({\scriptstyle{1\over 2}})
-{e^2B^2\over 48\pi^2T}\left(\ln {\mu\over 4\pi T}+\gamma_E\right)
\right]s
\nonumber \\
&&+{L^2 eB \over 2\pi} 
\left[
e^{-2a\sqrt{M^2+eB}}+{e^{-2a\sqrt{ 4\pi^2 T^2+M^2}}\over\sinh ({eBa\over \sqrt{ 4\pi^2 T^2+M^2}})}\right]s,
\label{z4}
\end{eqnarray}
where we took again the small $s$ limit and dropped a term independent of $s$. Last we add (\ref{zB3}), (\ref{ztBT3}), (\ref{zBa4}) 
and  (\ref{zBaT2}) to find the zeta function when $T\gg  a^{-1}\gg \sqrt{eB},M$
\begin{eqnarray}
\zeta(s)&=&{L^2a (eB)^{3\over 2} \over 2\pi}
\left[\sqrt{2}\zeta_H(-{\scriptstyle{1\over 2}},{\scriptstyle {M^2\over 2eB}})-
\zeta_H(-{\scriptstyle{1\over 2}},{\scriptstyle {M^2\over eB}})
\right]s+{L^2a} \left[
{\pi^2T^3\over 45}-{M^2T\over 12}
-{e^2B^2\over 48\pi^2T}\left(\ln {\mu\over 4\pi T}+\gamma_E\right)
\right]s
\nonumber \\
&&+{L^2  \over 8\pi } 
\left[{\zeta_R(3)\over a^2}+2M^2\ln \left(2\mu a\right)-{e^2B^2a^2\over18}\right]s+{L^2 eB \over 2\pi} 
{e^{-2a\sqrt{ 4\pi^2 T^2+M^2}}\over\sinh ({eBa\over \sqrt{ 4\pi^2 T^2+M^2}})}s,
\label{z41}
\end{eqnarray}
where we took the small $s$ limit and dropped two terms independent of $s$. 

To evaluate $\zeta(s)$ in the small plate distance limit, $a^{-1} \gg T, M, \sqrt{eB}$, we apply the Poisson resummation formula
to the $m$ sum in (\ref{zeta2}), and obtain
\begin{equation}
\zeta(s)={\beta\over 2}\left[\tilde{\zeta}_{M,B}(s)+\tilde{\zeta}_{M,B,a}(s)+\zeta_{M,B,T}(s) +
\tilde{\zeta}_{M,B,a,T}(s)\right],
\label{z2}
\end{equation}
%%%%%%%%%%%%%%%----PAGE 7----%%%%%%%%%%%%%%%%%%%%%%%%%%%%%%%%%%%%%
where 
\begin{equation}
\tilde{\zeta}_{M,B}(s)=
{L^2\mu^{2s} \over 4\pi^{3/2}\Gamma(s)}
\int_0^\infty dt \, t^{s-5/2} e^{-M^2t}{eBt \over \sinh eBt},
\label{ztB}
\end{equation}
\begin{equation}
\tilde{\zeta}_{M,B,a}(s)=
{L^2\mu^{2s} \over 4\pi^{3/2}\Gamma(s)}\sum_{n=1}^\infty 
\int_0^\infty dt \, t^{s-5/2}e^{-M^2t} {eBt \over \sinh eBt} e^{-{\pi^2n^2t}/a^2},
\label{ztBa}
\end{equation}
\begin{equation}
\zeta_{M,B,T}(s)=
{L^2 \mu^{2s}\over 2\pi^{3/2}\Gamma(s)}\sum_{m=1}^\infty 
\int_0^\infty dt \, t^{s-5/2}e^{-M^2t} {eBt \over \sinh eBt} e^{-{m^2\beta^2/4t}},
\label{zBT}
\end{equation}
\begin{equation}
\tilde{\zeta}_{M,B,a,T}(s)=
{L^2 \mu^{2s}\over 2\pi^{3/2}\Gamma(s)}\sum_{n=1}^\infty \sum_{m=1}^\infty 
\int_0^\infty dt \, t^{s-5/2} e^{-M^2t}{eBt \over \sinh eBt} e^{-(\pi^2n^2t/a^2+m^2\beta^2/4t)}.
\label{ztBaT}
\end{equation}
Comparing Eqs. (\ref{ztB}) - (\ref{ztBaT}) to (\ref{zB}) - (\ref{zBaT}), it is evident that $\tilde{\zeta}_{M,B}(s)=2\zeta_{M,B}(s)$, since
the part of $\zeta_{M,B}(s)$ proportional to $a^{-1}$ is negligible, that $\tilde{\zeta}_{M,B,a}(s)$ 
and $\tilde{\zeta}_{M,B,a,T}(s)$ are equal to $\tilde{\zeta}_{M,B,T}(s)$ and $\zeta_{M,B,a,T}(s)$ respectively, 
once we replace $a$ with $\beta/2$ and $\beta$ with $2a$, and that $\zeta_{M,B,T}(s)$ equals twice 
$\zeta_{M,B,a}(s)$
once we make the same replacement. Therefore we find that, in the small $s$ limit and when
$a^{-1} \gg M\gg T, \sqrt{eB}$, the zeta function is given by
\begin{equation}
\zeta(s)={L^2\beta} \left[
{\pi^2\over 720 a^3}-{M^2\over 48a}+{M^3 \over 6\pi}
-{e^2B^2a\over 48\pi^2}\left({\pi \over Ma}+\ln {\mu a\over 2\pi }+\gamma_E\right)
\right]s
+{ L^2 eB \over 2\pi }
\left[ 
{e^{-\beta M}\over\sinh ({eB\beta\over 2M})}
+{e^{-\beta\sqrt{ \pi^2 a^{-2}+M^2}}
\over\sinh \left({eB\beta\over 2\sqrt{ \pi^2 a^{-2} +M^2}}\right)}
\right]s,
\label{z5}
\end{equation}
when $a^{-1} \gg \sqrt{eB}\gg T, M$, the zeta function is
\begin{eqnarray}
\zeta(s)&=&{L^2\beta} \left[
{\pi^2\over 720 a^3}+{ (eB)^{3\over 2} \over 2\pi}(\sqrt 2 -1)
\zeta_R(-{\scriptstyle{1\over 2}})-{M^2\over 48 a}-{\sqrt{eB}M^2\over 4\pi}\left(1-{1\over\sqrt 2}\right)\zeta_R({\scriptstyle{1\over 2}})
-{e^2B^2a\over 48\pi^2}\left(\ln {\mu a\over 2\pi }+\gamma_E\right)
\right]s
\nonumber \\
&&+{L^2 eB \over2 \pi} 
\left[
2e^{-\beta\sqrt{eB+M^2}}+
{e^{-\beta\sqrt{ \pi^2 a^{-2}+M^2}}
\over\sinh \left({eB\beta\over 2\sqrt{ \pi^2 a^{-2} +M^2}}\right)}\right]s,
\label{z6}
\end{eqnarray}
and when $a^{-1} \gg T\gg  \sqrt{eB}, M$, the zeta function is given by
\begin{eqnarray}
\zeta(s)&=&{L^2\beta (eB)^{3\over 2} \over 2\pi}
\left[\sqrt{2}\zeta_H(-{\scriptstyle{1\over 2}},{\scriptstyle {M^2\over 2eB}})-
\zeta_H(-{\scriptstyle{1\over 2}},{\scriptstyle {M^2\over eB}})
\right]s+{L^2\beta} \left[
{\pi^2\over 720 a^3}-{M^2\over 48a}
-{e^2B^2a\over 48\pi^2}\left(\ln {\mu a\over 2\pi}+\gamma_E\right)
\right]s
\nonumber \\
&&+{L^2  \over \pi } 
\left[{\zeta_R(3)\over \beta^2}+{M^2\over 2}\ln \left(\mu\beta \right) -{e^2B^2\beta^2\over 288}\right]s+{L^2 eB \over 2\pi} 
{e^{-\beta\sqrt{ \pi^2 a^{-2}+M^2}}
\over\sinh \left({eB\beta\over 2\sqrt{ \pi^2 a^{-2} +M^2}}\right)}s.
\label{z6a}
\end{eqnarray}

Next  we evaluate $\zeta(s)$ in the strong magnetic field limit, $\sqrt{eB}\gg T, a^{-1}, M$, 
and apply the Poisson resummation formula to both the $n$ and $m$ sums in (\ref{zeta2}), to find
\begin{equation}
\zeta(s)=a\beta[ \zeta_{W}(s)+\tilde\zeta(s)]
\label{z7}
\end{equation}
where
\begin{equation}
\zeta_{W}(s)= {L^2 \mu^{2s}\over 16\pi^2\Gamma(s)}
\int_0^\infty dt \, t^{s-3}\left(1+a^{-1}\sqrt{\pi t}\right)e^{-M^2t} {eBt \over \sinh eBt},
\label{zW}
\end{equation}
and 
\begin{equation}
\tilde\zeta(s)= {L^2 \mu^{2s}\over 16\pi^2\Gamma(s)}
\int_0^\infty dt \, t^{s-3}e^{-M^2t} {eBt \over \sinh eBt}\left(\sum_{n,m=-\infty}^\infty e^{-{a^2}n^2 /t} e^{-{ \beta^2}m^2/4 t}-1\right).
\label{z8}
\end{equation}
%%%%%%%%%%%%%%%----PAGE 8----%%%%%%%%%%%%%%%%%%%%%%%%%%%%%%%%%%%%%
Eq. (\ref{zW}), once we neglect the term proportional to $a^{-1}$,
yields the zeta function of the  one-loop Weisskopf effective Lagrangian for massive scalar QED \cite{Weisskopf:1996bu}.
In the strong magnetic field limit, we set $e^{-M^2t}\simeq 1-M^2t$, integrate, and obtain
$$
\zeta_{W}(s)={L^2 (eB)^2\over 8\pi^2\Gamma(s)}\left({\mu^2\over eB}\right)^s\left[
{(1-2^{1-s})}\Gamma(s-1)\zeta_R(s-1)
-(1-2^{-s}){M^2\over eB}\Gamma(s)\zeta_R(s)
\right],
$$
and, for small $s$, we find
\begin{equation}
\zeta_{W}(s)={L^2 (eB)^2\over 96\pi^2}\left(\ln{eB\over 3\mu^2}-{1\over 2} -s^{-1}
+{M^2\over eB}6\ln 2
\right)s,
\label{zW2}
\end{equation}
where we used the interesting numerical fact \cite{Erdas:2009zh,Erdas:2010yq}
$${6\over \pi^2}\zeta'_R(2)-\log \pi-\gamma_E=-2.2918\approx-\ln 6 -{1\over 2} .$$
Once we compare Eq. (\ref{zW2}) to the well known result for the Weisskopf Lagrangian \cite{Weisskopf:1996bu,Erdas:2010yq}, we realize that we must take the 
arbitrary parameter $\mu=M$, and we'll do that in all our results containing $\mu$.
We evaluate  $\tilde\zeta(s)$ by using 
$$
{1\over\sinh eBt}\approx 2e^{-eBt},
$$and changing integration variable from $t$ to
$\sqrt{n^2 a^2+m^2\beta^2/4 \over {eB+M^2}}t$ in Eq. (\ref{z8}), to find
$$
\tilde\zeta(s)= {L^2\mu^{2s} eB\over 8\pi^2\Gamma(s)} \sum_{n,m=-\infty}^\infty
\left({n^2 a^2+m^2\beta^2/4 \over {eB+M^2}}\right)^{{s-1}\over 2}\int_0^\infty dt \, t^{s-2} 
e^{-{1\over 2}(t+t^{-1})\sqrt{(eB+M^2)(4n^2 a^2+m^2\beta^2 )}},
$$where the term with $m=n=0$ is excluded and only terms with $n=0,\pm 1$ and $m=0,\pm 1$ contribute significantly to the double sum. 
We integrate using the saddle point method and, using (\ref{z7}) and  (\ref{zW2}), obtain
the zeta function in the strong magnetic field and small $s$ limit,
\begin{equation}
\zeta(s)={L^2a (eB)^2\over 96\pi^2T}\left(\ln{eB\over 3M^2}-{1\over 2}
\right)s+{L^2 a eB(eB+M^2)^{1\over 4}  \over 4\pi^{3\over 2} T}
\left[{e^{-2a\sqrt{eB+M^2}}\over a^{3\over 2}}+{2^{3\over 2}e^{-\beta\sqrt{eB+M^2}}\over \beta^{3\over 2}}+
{2e^{-\sqrt{(eB+M^2)(4a^2+\beta^2)}}\over \left({a^2+{\beta^2\over 4} }\right)^{3\over 4}}\right]s,
\label{z10}
\end{equation}
where we neglected higher order terms in $M^2\over eB$ and terms that do not depend on $s$.

Last we use (\ref{z7}) to evaluate $\zeta(s)$ in the large mass limit, $M\gg T, a^{-1}, \sqrt{eB}$. We first obtain $\zeta_{W}(s)$
for $M\gg \sqrt{eB}$ and small $s$
$$
\zeta_{W}(s)={L^2 M^4\over 16\pi^2}\left({3\over 4} +\ln{\mu\over M}+{1\over 2s}
-{e^2B^2\over 6M^4}
\right)s.
$$
To evaluate $\tilde\zeta(s)$, we change the  integration variable from $t$ to
$\sqrt{n^2 a^2+m^2\beta^2/4 \over M}t$ in Eq. (\ref{z8}), retain only terms with $n=0,\pm 1$ and $m=0,\pm 1$ in the double sum, use the saddle point method to integrate and, for small $s$, obtain
\begin{equation}
\zeta(s)={3L^2a\beta M^4\over 64 \pi^2}s+{L^2 a\beta eB\sqrt{M} \over 8\pi^{3/2} }
\left[{e^{-2aM}\over a^{3/2}\sinh\left({eBa\over M}\right)}+{2^{3/2}e^{-\beta M}\over \beta^{3/2}\sinh\left({eB\beta\over 2M}\right)}+
{2e^{-M\sqrt{4a^2+\beta^2 }}\over \left({a^2+{\beta^2\over 4} }\right)^{3/4}\sinh\left({eB\over 2M}\sqrt{4a^2+{\beta^2} }\right)}\right]s,
\label{z11}
\end{equation}
where we set $\mu=M$, and neglected higher order terms in $eB\over M^2$ and terms that do not depend on $s$.
%%%%%%%%%%%%%%%%%%%%%%%%%%%%%%%%%%%%%%%%%%%%%%%%%%%%%%%%%%%
\section{ Free energy and Casimir pressure}
\label{3}
We use Eq. (\ref{Fandzeta}) to calculate the free energy and are able take the derivative of the zeta function easily, by using the
fact  that the derivative of $G(s)/\Gamma(s)$ at $s=0$ is simply $G(0)$, if $G(s)$ is a well behaved function.  
Using (\ref{zeta2}) and
%%%%%%%%%%%%%%%----PAGE 9----%%%%%%%%%%%%%%%%%%%%%%%%%%%%%%%%%%%%%
our other results for the zeta function (\ref{z}), (\ref{z2}),  and (\ref{z7}), we are able to write four expressions of 
the free energy, all equivalent to each other, 
\begin{equation}
F=-{L^2 \over 4\pi\beta}
\int_0^\infty dt \, t^{-2} e^{-M^2t}{eBt \over \sinh eBt}\left(\sum_{n=0}^\infty e^{-{\pi^2\over a^2}n^2 t}\right)
\left(\sum_{m=-\infty}^\infty e^{-{4\pi^2\over \beta^2}m^2 t}\right),
\label{F2}
\end{equation}
\begin{equation}
F=-
{L^2a \over 8\pi^{3/2}\beta}
\int_0^\infty dt \, t^{-5/2}e^{-M^2t} {eBt \over \sinh eBt}\left({\sqrt{\pi t}\over a}+\sum_{n=-\infty}^\infty e^{-{n^2a^2\over t}}\right)
\left(\sum_{m=-\infty}^\infty e^{-{4\pi^2\over \beta^2}m^2 t}\right)
\label{F2a}
\end{equation}
better suited for a high temperature expansion ($2Ta\gg 1$, $2T\gg \sqrt{eB}/\pi$ and $T\gg M$),
\begin{equation}
F=-
{L^2 \over 8\pi^{3/2}}
\int_0^\infty dt \, t^{-5/2}e^{-M^2t} {eBt \over \sinh eBt}\left(\sum_{n=0}^\infty e^{-{\pi^2\over a^2}n^2 t}\right)
\left(\sum_{m=-\infty}^\infty e^{-{m^2 \beta^2\over 4t} }\right)
\label{F2b}
\end{equation}
better suited for a small plate distance expansion ($2Ta\ll 1$, $a^{-1}\gg \sqrt{eB}/\pi$ and $a^{-1}\gg M$), and
\begin{equation}
F=-
{L^2a \over 16\pi^{2}}
\int_0^\infty dt \, t^{-3}e^{-M^2t} {eBt \over \sinh eBt}\left({\sqrt{\pi t}\over a}+\sum_{n=-\infty}^\infty e^{-{n^2a^2\over t}}\right)
\left(\sum_{m=-\infty}^\infty e^{-{m^2 \beta^2\over 4t} }\right)
\label{F2c}
\end{equation}
better suited for strong magnetic field or large mass expansion. The last equation has been obtained by other authors
\cite{CougoPinto:1998jg} in a very similar form, and used by them to write the free energy as an infinite sum of
modified Bessel functions.

None of the four expressions, (\ref{F2}) - (\ref{F2c}), can be evaluated in closed form for arbitrary values of the four
quantities $M$, $B$, $a$ and $T$, but it is possible to use them to evaluate
numerically the free energy for any values of these four relevant quantities. However,
using our results from Sec. \ref{2}, we found simple analytic expressions for the free energy when one or some of 
those four quantities are small or large. 
To obtain the free energy in the high temperature limit, we use 
(\ref{z3}) -  (\ref{z41}) and find  
\begin{equation}
F=-V \left[
{\pi^2T^4\over 45}-{M^2T^2\over 12}+{M^3T \over 6\pi}
-{e^2B^2\over 48\pi^2}\left({\pi T\over M}+\ln {M\over 4\pi T}+\gamma_E\right)
\right]
-{ L^2T eB \over 2\pi }\left[ 
{e^{-2aM}\over 2\sinh ({eBa\over M})}+
{e^{-2a\sqrt{ 4\pi^2 T^2+M^2}}\over\sinh ({eBa\over \sqrt{ 4\pi^2 T^2+M^2}})}
\right],
\label{F3}
\end{equation}
valid for $T\gg M\gg \sqrt{eB}, a^{-1}$ and where $V=L^2a$ is the volume of the slab,
\begin{eqnarray}
F&=&-V\left[
{\pi^2T^4\over 45}+{ (eB)^{3\over 2}T \over 2\pi}(\sqrt 2 -1)
\zeta_R(-{\scriptstyle{1\over 2}})-{M^2T^2\over 12}-{\sqrt{eB}M^2T\over 4\pi}\left(1-{1\over\sqrt 2}\right)\zeta_R({\scriptstyle{1\over 2}})
-{e^2B^2\over 48\pi^2}\left(\ln {M\over 4\pi T}+\gamma_E\right)
\right]
\nonumber \\
&&-{L^2 T eB \over 2\pi} 
\left[
e^{-2a\sqrt{M^2+eB}}+{e^{-2a\sqrt{ 4\pi^2 T^2+M^2}}\over\sinh ({eBa\over \sqrt{ 4\pi^2 T^2+M^2}})}\right],
\label{F4}
\end{eqnarray}
valid for $T\gg \sqrt{eB}\gg M, a^{-1}$, and
\begin{eqnarray}
F&=&-V \left[
{\pi^2T^4\over 45}-{M^2T^2\over 12}
-{e^2B^2\over 48\pi^2}\left(\ln {M\over 4\pi T}+\gamma_E\right)
\right]-{V T (eB)^{3\over 2} \over 2\pi}
\left[\sqrt{2}\zeta_H(-{\scriptstyle{1\over 2}},{\scriptstyle {M^2\over 2eB}})-
\zeta_H(-{\scriptstyle{1\over 2}},{\scriptstyle {M^2\over eB}})
\right]
\nonumber \\
&&-{L^2 T \over 8\pi } 
\left[{\zeta_R(3)\over a^2}+2M^2\ln \left(2M a\right)-{e^2B^2a^2\over18}\right]-{L^2T eB \over 2\pi} 
{e^{-2a\sqrt{ 4\pi^2 T^2+M^2}}\over\sinh ({eBa\over \sqrt{ 4\pi^2 T^2+M^2}})},
\label{F41}
\end{eqnarray}
valid when $T\gg  a^{-1}\gg \sqrt{eB},M$. Notice that Eqs. (\ref{F4}) and  (\ref{F41}), once we set $M=0$, are in full agreement with the
results of Ref. \cite{Erdas:2013jga}, where we studied the Casimir effect
for a massless scalar field. In addition  to reproducing the results of our previous paper, Eqs. (\ref{F4}) and  (\ref{F41}) contain the corrections 
to those results due to a ``light'' scalar mass. Eq. (\ref{F3}) gives the free energy in the case of an ``intermediate'' scalar mass, since it is valid when
$M$ is much smaller than the temperature and much larger than the inverse plate distance and $\sqrt{eB}$.
Eqs. (\ref{F3}) - (\ref{F41}) show that, 
in the high temperature limit, the dominant term in the free energy is the Stefan-Boltzmann term $-{\pi^{2}\over 45}VT^4$, as expected. 
Terms in the free energy that have a linear dependence on $a$, such as this one, are 
uniform energy density term. If the medium outside the plates is at the same temperature $T$ and has the same magnetic field present
as the medium 
%%%%%%%%%%%%%%%----PAGE 10----%%%%%%%%%%%%%%%%%%%%%%%%%%%%%%%%%%%%%
between the plates, uniform energy density terms do not
contribute to the Casimir pressure. Only if there is not a magnetic field and the temperature is zero outside the plates, uniform energy
density terms contribute a constant pressure. In this
work we assume that the same magnetic field is present between and outside 
the plates, and that the medium outside the plates is at the same temperature as the one between the
plates, therefore we will neglect contributions to the Casimir pressure from uniform energy density terms.

The pressure $P$ on the plates is
$$
P=-{1\over L^2}{\partial F\over\partial a},
$$
and therefore, for $T\gg M\gg \sqrt{eB}, a^{-1}$
\begin{equation}
P=-{ T eB \over \pi }\left[ {M\over 2}
{ e^{-2aM}\over \sinh ({eBa\over M})}+
\sqrt{ 4\pi^2 T^2+M^2}{e^{-2a\sqrt{ 4\pi^2 T^2+M^2}}\over\sinh ({eBa\over \sqrt{ 4\pi^2 T^2+M^2}})}
\right],
\label{P2}
\end{equation}
for $T\gg  \sqrt{eB}\gg M, a^{-1}$
\begin{equation}
P=-{ T eB \over \pi }\left[ 
\sqrt{M^2+eB}e^{-2a\sqrt{M^2+eB}}+
\sqrt{ 4\pi^2 T^2+M^2}{e^{-2a\sqrt{ 4\pi^2 T^2+M^2}}\over\sinh ({eBa\over \sqrt{ 4\pi^2 T^2+M^2}})}
\right],
\label{P3}
\end{equation}
and for $T\gg  a^{-1}\gg M, \sqrt{eB}$
\begin{equation}
P=
-{ T \over 4\pi } 
\left[{\zeta_R(3)\over a^3}-{M^2\over a}+{e^2B^2a\over18}\right]
-{ T eB \over \pi }\sqrt{ 4\pi^2 T^2+M^2}{e^{-2a\sqrt{ 4\pi^2 T^2+M^2}}\over\sinh ({eBa\over \sqrt{ 4\pi^2 T^2+M^2}})}.
\label{P4}
\end{equation}
Notice that, in Eqs. (\ref{P2}) - (\ref{P4}), we left out some terms that are negligibly small.

We use (\ref{z5}) -  (\ref{z6a}) to obtain the free energy in the small plate distance limit
\begin{equation}
F=-{L^2} \left[
{\pi^2\over 720 a^3}-{M^2\over 48a}+{M^3 \over 6\pi}
-{e^2B^2a\over 48\pi^2}\left({\pi \over Ma}+\ln {M a\over 2\pi }+\gamma_E\right)
\right]
-{ L^2 TeB \over 2\pi }
\left[ 
{e^{-\beta M}\over\sinh ({eB\beta\over 2M})}
+{e^{-\beta\sqrt{ \pi^2 a^{-2}+M^2}}
\over\sinh \left({eB\beta\over 2\sqrt{ \pi^2 a^{-2} +M^2}}\right)}
\right]\label{F5}
\end{equation}
for $a^{-1} \gg M\gg T, \sqrt{eB}$,
\begin{eqnarray}
F&=&-{L^2} \left[
{\pi^2\over 720 a^3}+{ (eB)^{3\over 2} \over 2\pi}(\sqrt 2 -1)
\zeta_R(-{\scriptstyle{1\over 2}})-{M^2\over 48 a}-{\sqrt{eB}M^2\over 4\pi}\left(1-{1\over\sqrt 2}\right)\zeta_R({\scriptstyle{1\over 2}})
-{e^2B^2a\over 48\pi^2}\left(\ln {M a\over 2\pi }+\gamma_E\right)
\right]
\nonumber \\
&&-{L^2 T eB \over2 \pi} 
\left[
2e^{-\beta\sqrt{eB+M^2}}+
{e^{-\beta\sqrt{ \pi^2 a^{-2}+M^2}}
\over\sinh \left({eB\beta\over 2\sqrt{ \pi^2 a^{-2} +M^2}}\right)}\right]
\label{F6}
\end{eqnarray}
for  $a^{-1} \gg \sqrt{eB}\gg T, M$, and
\begin{eqnarray}
F&=&-{L^2 (eB)^{3\over 2} \over 2\pi}
\left[\sqrt{2}\zeta_H(-{\scriptstyle{1\over 2}},{\scriptstyle {M^2\over 2eB}})-
\zeta_H(-{\scriptstyle{1\over 2}},{\scriptstyle {M^2\over eB}})
\right]-{L^2} \left[
{\pi^2\over 720 a^3}-{M^2\over 48a}
-{e^2B^2a\over 48\pi^2}\left(\ln {M a\over 2\pi}+\gamma_E\right)
\right]
\nonumber \\
&&-{L^2T  \over \pi } 
\left[{\zeta_R(3)T^2}+{M^2\over 2}\ln \left(M\over T\right) -{e^2B^2\over 288T^2}\right]-{L^2 TeB \over 2\pi} 
{e^{-\beta\sqrt{ \pi^2 a^{-2}+M^2}}
\over\sinh \left({eB\beta\over 2\sqrt{ \pi^2 a^{-2} +M^2}}\right)}
\label{F61}
\end{eqnarray}
 when $a^{-1} \gg T\gg  \sqrt{eB}, M$. Eqs. (\ref{F6}) and  (\ref{F61}) contain the corrections 
to the results of Ref. \cite{Erdas:2013jga} due to a ``light'' scalar mass, and Eq. (\ref{F5}) shows the free energy in the case of an ``intermediate'' scalar mass.
The dominant term in (\ref{F5}) -  (\ref{F61}) is $-{\pi^{2}\over 720}{L^2\over a^3}$, which is the familiar vacuum Casimir energy 
for a complex scalar field, and for the photon field \cite{Casimir:1948dh}.
The Casimir pressure is the same in all three cases of small plate distance
\begin{equation}
P=-
{\pi^{2}\over 240 a^4}+
{\pi eB \over 2a^3\sqrt{ \pi^2 a^{-2}+M^2}} {e^{-\beta\sqrt{ \pi^2 a^{-2}+M^2}}
\over\sinh \left({eB\beta\over 2\sqrt{ \pi^2 a^{-2} +M^2}}\right)}
-{e^2B^2 \over 48\pi^{2}}( \ln{Ma\over 2\pi}+1),
\label{P6}
\end{equation}
%%%%%%%%%%%%%%%----PAGE 11----%%%%%%%%%%%%%%%%%%%%%%%%%%%%%%%%%%%%%
and, again, we left out some negligibly small terms.

For strong magnetic field, $\sqrt{eB}\gg T, a^{-1}, M$, the free energy is found using (\ref{z10})
\begin{equation}
F=-{V (eB)^2\over 96\pi^2}\left(\ln{eB\over 3M^2}-{1\over 2}
\right)-{L^2 eB(eB+M^2)^{1/4}  \over 4\pi^{3/2} }
\left[{e^{-2a\sqrt{eB+M^2}}\over \sqrt{a}}+{2^{3/2}a e^{-\beta\sqrt{eB+M^2}}\over \beta^{3/2}}+
{2a e^{-\sqrt{(eB+M^2)(4a^2+\beta^2)}}\over \left({a^2+{\beta^2\over 4} }\right)^{3/4}}\right],
\label{F7}
\end{equation}
where the dominant term is the one-loop vacuum effective potential for scalar QED \cite{Erdas:2010yq}, 
proportional to the volume of the slab. If we set $M=0$ in  (\ref{F7}), we reproduce the result of Ref. \cite{Erdas:2013jga}
obtained for a massless scalar field in the presence of a strong magnetic field.
The effective potential is a uniform energy density term and 
therefore, under our assumptions, does not contribute to the Casimir pressure.
The pressure, in the strong magnetic field case, is given by
\begin{equation}
P=-{eB(eB+M^2)^{3/4}  \over 2\pi^{3/2} }
\left[{e^{-2a\sqrt{eB+M^2}}\over \sqrt{a}}+
{2a^2 e^{-\sqrt{(eB+M^2)(4a^2+\beta^2)}}\over \left({a^2+{\beta^2\over 4} }\right)^{5/4}}\right],
\label{P8}
\end{equation}
where we neglected uniform energy density terms and some smaller terms.

Last we examine the large mass limit, $M\gg T, a^{-1}, \sqrt{eB}$ and, using (\ref{z11}), we obtain the free
energy in this limit
\begin{equation}
F=-{3V M^4\over 64 \pi^2}-{L^2  eB\sqrt{M} \over 8\pi^{3/2} }
\left[{e^{-2aM}\over \sqrt{a}\sinh\left({eBa\over M}\right)}+{2^{3/2}ae^{-\beta M}\over \beta^{3/2}\sinh\left({eB\beta\over 2M}\right)}+
{2ae^{-M\sqrt{4a^2+\beta^2}}\over \left({a^2+{\beta^2\over 4} }\right)^{3/4}\sinh\left({eB\over 2M}\sqrt{4a^2+\beta^2 }\right)}\right].
\label{F71}
\end{equation}
Also in this case the dominant term is the scalar QED effective potential and it will not contribute to the
pressure, since it is a uniform energy density term. The Casimir pressure in the large mass limit is given by
\begin{equation}
P=-{ eB M^{3/2} \over 4\pi^{3/2} }
\left[{e^{-2aM}\over \sqrt{a}\sinh\left({eBa\over M}\right)}+
{2a^2e^{-M\sqrt{4a^2+\beta^2}}\over \left({a^2+{\beta^2\over 4} }\right)^{5/4}\sinh\left({eB\over 2 M}\sqrt{4a^2+{\beta^2} }\right)}\right].
\label{P9}
\end{equation}
%%%%%%%%%%%%%%%%%%%%%%%%%%%%%%%%%%%%%%%%%%%%%%%%%%%%%%
\section{Discussion and conclusions}
\label{4}
In this work we used the zeta function method to investigate the finite
temperature Casimir effect
of a charged, massive scalar field confined between two perfectly conducting parallel plates and in the 
presence of a uniform magnetic field. We derived four expressions 
of the zeta function (\ref{zeta2}), (\ref{z}), (\ref{z2}),  and (\ref{z7}),  which are exact to all orders in $B$, $a$, $M$, and $T$, and used them to obtain expressions for the free energy of the scalar field and for the Casimir pressure on the plates in the case of high temperature
($T \gg a^{-1}, \sqrt{eB}, M$), small plate distance ($a^{-1} \gg T, \sqrt{eB}, M$), strong magnetic field
($\sqrt{eB}\gg T, a^{-1}, M$),  and large mass ($M\gg T, a^{-1}, \sqrt{eB}$).

We numerically evaluated the free energy with very high precision, using three of the exact expressions 
we obtained,  and we compared the exact numerical values of the free energy to the values obtained from
our simple analytic expressions. In the high temperature case we found that, for $T/2 \ge a^{-1}, M, \sqrt{eB}$,
Eq. (\ref{F3}) is within $1.3\%$ of the exact value of the free energy when $M\ge a^{-1}, \sqrt{eB}$,  
Eq. (\ref{F4}) is within $2.9\%$ of the exact value of the free energy when $\sqrt{eB}\ge a^{-1}, M$, and
Eq. (\ref{F41}) is within $4.2\%$ of the exact value of the free energy when $a^{-1}\ge \sqrt{eB}, M$.
For $T/4 \ge a^{-1}, M, \sqrt{eB}$, Eqs. (\ref{F3}) -  (\ref{F41}) are within $0.5\%$ or less of the exact value of the free energy, showing a very 
rapid convergence of Eqs. (\ref{F3}) - (\ref{F41}) to the exact values of the free energy.
These three equations are a simple analytic expression 
of $F$ in the high temperature limit, valid for all values of  $B$, $a$, and $M$, and have a discrepancy  from 
the exact value of $F$ that is not larger than $4.2\%$, as long as $T/2 \ge a^{-1}, M, \sqrt{eB}$. An equally accurate expression of the Casimir pressure, valid for 
$T/2 \ge a^{-1}, M, \sqrt{eB}$, is obtained immediately from (\ref{F3}) - (\ref{F41}), and is shown in (\ref{P2}) - (\ref{P4}). 

When investigating the small plate distance limit we found that, for $a^{-1}/4\ge 2T,M,\sqrt{eB}$,
Eq. (\ref{F5}) is within $2.9\%$ of the exact value of the free energy when $M\ge 2T,\sqrt{eB}$,  
Eq. (\ref{F6}) is within $5.3\%$ of the exact value of the free energy 
when $\sqrt{eB}\ge 2T,M$, and Eq. (\ref{F61}) is within $8.2\%$ of the exact value of the free energy 
when $2T\ge M,\sqrt{eB}$.
For $a^{-1}/8\ge 2T,M,\sqrt{eB}$,
Eqs. (\ref{F5}) - (\ref{F61}) are within $1\%$ or less of the exact value of the free energy, showing once more
a rapid convergence of our analytical
expressions to the exact value of the free energy. Eqs. (\ref{F5}) - (\ref{F61})
are a simple analytic expression
of the free energy in the small plate distance limit, valid for all values of 
$B$, $M$ and $T$, and with
%%%%%%%%%%%%%%%----PAGE 12----%%%%%%%%%%%%%%%%%%%%%%%%%%%%%%%%%%%%%
a discrepancy of no more than $8.2\%$ from 
the exact value of $F$ when $a^{-1}/4\ge 2T,M,\sqrt{eB}$. The pressure in the case of small plate distance, obtained immediately from
 (\ref{F5}) - (\ref{F61}) and shown in  (\ref{P6}), is similarly accurate.

We find that, in the case of strong magnetic field or large mass, Eqs. (\ref{F7}) and (\ref{F71}) are even more accurate. 
For $\sqrt{eB}/2\ge 2T,M,a^{-1}$, Eq. (\ref{F7}) is within $4.6\%$ of the exact value of the free energy, and for
$\sqrt{eB}/4\ge 2T,M,a^{-1}$, Eq. (\ref{F7}) is within $0.0001\%$ of the exact value of the free energy, showing an extremely
rapid convergence to the exact value of $F$. The large mass limit of Eq. (\ref{F71}), for $M/2\ge 2T,\sqrt{eB},a^{-1}$,  
is within $0.05\%$ of the exact value of the free energy.

If we set $T=0$ in Eqs. (\ref{F5}) and (\ref{F6}), and eliminate terms that do not depend on $a$ and
uniform energy density terms, 
we obtain the same quantity, which is the Casimir energy $E_C$ due to a massive and charged
scalar field in a magnetic field in the limit of small plate distance ($a^{-1}\gg \sqrt{eB}, M$)
\begin{equation}
{E_C \over L^2}=
-{\pi^2\over 720 a^3}+{M^2\over 48a}
+{e^2B^2a\over 48\pi^2}\ln {M a\over 2\pi }.
\label{EC1}
\end{equation}
If we do the same in Eqs. (\ref{F7}) and (\ref{F71}), we find that, for strong magnetic field ($\sqrt{eB}\gg a^{-1}, M$)
\begin{equation}
{E_C \over L^2}=
-{ eB(eB+M^2)^{1/4}  \over 4\pi^{3/2} \sqrt{a}}
e^{-2a\sqrt{eB+M^2}},
\label{EC2}
\end{equation}
and for large scalar mass ($M\gg a^{-1}, \sqrt{eB}$)
\begin{equation}
{E_C \over L^2}=
-{eB\sqrt{M} \over 8\pi^{3/2} }
{e^{-2aM}\over \sqrt{a}\sinh\left({eBa\over M}\right)}.
\label{EC3}
\end{equation}
Eqs. (\ref{EC1}) - (\ref{EC3}) show that scalar mass, as it grows, inhibits the Casimir energy. The situation is different for a growing magnetic field which,
in the case of small plate distance, boosts the Casimir energy, as we can see from Eq.
(\ref{EC1}) where $\ln {M a\over 2\pi }$ is negative because $aM\ll 1$. In the case of  strong magnetic field and large scalar mass, Eqs. 
(\ref{EC2}) and (\ref{EC3}) show that magnetic field, as it grows, inhibits the Casimir energy  
 as it is also shown in \cite{CougoPinto:1998td}
and, in the case of a massless scalar, in \cite{Erdas:2013jga}. 
Our results, simple analytic expressions for $E_C$, are more explicit than those of \cite{CougoPinto:1998td} where the magnetic field correction
to the Casimir energy is presented as an infinite sum of integrals, and more general than those of \cite{Erdas:2013jga}, where only the case of a
massless scalar field is examined. Notice that Eq. (\ref{EC2}), once we set $M=0$, agrees with the result of Ref. \cite{Erdas:2013jga} and 
agrees with \cite{CougoPinto:1998td} on the dependence of $E_C$ from $a$ and $B$, but disagrees with this paper
for the overall sign.
%%%%%%%%%%%%%%%%%%%%%%%%%%%%%%%%%%%%%%%%%%%%%%%%%%%%%%%%%%%%%%%%%%%

%%%%%%%%%%%%%%%%%%%%%%%%%%%%%%%%%%%%%%%%%%%%%%%%%%%%%%%%%%%%%%%%%%%
\end{document}